\documentclass[aps,preprint,prd,nofootinbib]{revtex4}
\usepackage{graphicx}
\usepackage{psfrag}

\begin{document}

\title{The Production of $X(3940)$ and $X(4160)$ in $B_c$ decays}
\author{Zhi-Hui Wang$^{[1],[2]}$\footnote{2013086@nun.edu.cn}, Yi Zhang$^{[1]}$, Tian-hong Wang$^{[2]}$, Yue Jiang$^{[2]}$, Guo-Li Wang$^{[2]}$\\}
\address{$^1$School of Electrical and Information Engineering, Beifang University of Nationalities, Yinchuan, 750021,\\
$^2$Department of Physics, Harbin Institute of
Technology, Harbin, 150001}

 \baselineskip=20pt

\begin{abstract}
Considering $X(3940)$ and $X(4160)$ as $\eta_c(3S)$ and $\eta_c(4S)$,
we study the productions of $X(3940)$ and $X(4160)$
in exclusive weak decays of $B_c$ meson by the improved Bethe-Salpeter(B-S) Method.
Using the relativistic B-S equation and Mandelstam formalism,
we calculate the corresponding decay form factors.
The predictions of the corresponding
branching ratios are:
$Br(B_c^+\to X(3940)e^+\nu_e)$$=1.0\times10^{-4}$ and $Br(B_c^+\to X(4160)e^+\nu_e)=2.4\times10^{-5}$.
That will provide us a new way to observe the $X(3940)$ and $X(4160)$ in the future,
as well as to improve the knowledge of $B_c$ meson decay.
 \vspace*{0.5cm}


\end{abstract}

\maketitle
\section{Introduction}

In the last few years, lots of new particles have been observed,
such as charmomium-like state was first observed in $e^+e^-$ annihilation by Belle in 2002~\cite{FirstCharm}.
Then many more charmomium-like states were discovered in experiments,
and the family of charmonium-like states have become very abundant.
Such as the $X(3872)$ resonance was discovered by Belle Collaboration
 through the channel $B^{\pm}\to K^{\pm}\pi^+\pi^-J/ \psi$~\cite{3872}.
The $X(3915)$ was reported by Belle Collaboration
in $\gamma\gamma\to\omega J/\psi$~\cite{3915}.
$X(3940)$ was observed from the inclusive process
$e^+e^-\to J/\psi X(3940)$, with the mass $(3943\pm6\pm6)$ MeV~\cite{3940}.
Latter Belle Collaboration confirmed $X(3940)$ by the process
$e^+e^-\to J/\psi D^*\bar D^*$,
and they also reported a new charmonium-like state $X(4160)$~\cite{39404160}.
Now the Particle Data Group(PDG) give the mass and width of $X(3940)$: $M=(3942^{+7}_{-6}\pm6)$ MeV,
$\Gamma=(37^{+26}_{-15}\pm8)$ MeV,
the mass and width of $X(4160)$: $M=(4156^{+25}_{-20}\pm15)$ MeV, $\Gamma=(139^{+111}_{-61}\pm21)$ MeV~\cite{PDG}.

The charmomium-like states provide us a good way to study the
nonperturbative behavior of QCD, so they have attracted a lot of attention of theorists and experimentalists.
People had detailed summarized the present experimental
status of the $XYZ$ particles,
gave the productions and properties of $XYZ$ states~\cite{liuxiang1,zhushilin1,stephen1,zhao1,kim}.
As two of these new observed particles, there are already some theoretical studies on $X(3940)$ and $X(4160)$.
Ref.~\cite{kim} calculate the mass of $X(3940)$ as $J^{PC}=2^{++}$.
Ref.~\cite{Lu} gave the production of $X(3940)$ which was assumed as $3^1S_0$ state
in weak decay of $B_c$ in the framework of the light-cone QCD sum rules approach.
Ref.~\cite{39401} had studied the inclusive production of $X(3940)$
in the decay of ground bottomnium state $\eta_b$ by the NRQCD factorization formula,
and they also considered $X(3940)$ as the excited $\eta_c(3S)$ state.
Using the NRQCD factorization approach, Ref.~\cite{zhuruilin} calculated the branching fractions of
$\Upsilon(nS)\to J/\psi+X$ with $X=X(3940)$ or $X=X(4160)$.
In Ref.~\cite{liuxiang2}, they also explored the properties and strong decays of $X(3940)$ and $X(4160)$
as the $\eta_c(3S)$ and $\eta_c(4S)$, respectively.
Ref.~\cite{41601} calculated the strong decay of $X(4160)$ which was assumed as $\chi_{c0}(3P)$,
$\chi_{c1}(3P)$, $\eta_{c2}(2D)$ or $\eta_c(4S)$ by the $^3P_0$ model.
Ref.~\cite{th6} calculated the strong decays of $\eta_c(nS)$,
they found that the explanation of $X(3940)$
as $\eta_c(3S)$ is possible and the assignment of $X(4160)$ as $\eta_c(4S)$ can not
be excluded.
According to the strong decay of $X(3940)$ and $X(4160)$ in Ref.~\cite{liuxiang2,41601,th6},
considering $X(3940)$ and $X(4160)$ as $\eta_c(3S)$ and $\eta_c(4S)$($J^{PC}=0^{-+}$) are possible.

In this paper we will consider the possibilities of $X(3940)$ and $X(4160)$
as radial high excited states $\eta_c(3S)$ and $\eta_c(4S)$, respectively.
We focus on the productions of $X(3940)$ and $X(4160)$
in exclusive weak decays of $B_c$ meson by the improved the Bethe-Salpeter(B-S) Method.
On the one hand, the higher excited states have larger relativistic correction than
the corresponding ground state, a relativistic model is needed in a careful study.
On the other hand, this study can improve the knowledge of $B_c$ meson,
and $B_c$ meson only decay weakly which is an ideal particle to study the weak decays.
In recent years, more and more people had studied the $B_c$ meson by different methods,
such as different relativistic constituent quark models~\cite{bc1,bc2,bc3,bc4,bc5,Ebert11,Ebert12,Ebert13,Ebert14,Ebert15,Ivanov1,Ivanov2,Ivanov3,Ivanov4},
the covariant light-front quark model~\cite{bc6,bc7} and perturbative QCD factorization approach~\cite{bc8}.
In these literatures, they studied the nature of $B_c$ meson
by the semileptonic and nonleptonic decays of $B_c$, the $CP$ violation in two-body hadronic decays of $B_c$,
rare semileptonic decays of $B_c$ etc.
We also discussed the properties of $B_c$ meson by the improved B-S method,
include $B_c$ decays to $P-$wave mesons,
 the rare weak decays and rare radiative decays of $B_c$, the nonleptonic charmless decays of $B_c$,
 and so on~\cite{bc-pwave,heavy-light,bc9,bc10,bc11,bc12,bc13}.
 In previous papers, we focused on $B_c$ decays to $1S,~2S,~1P,$ and $2P$ states,
 because when the final states were $3S,~4S$ states, the corresponding branching ratios were very small,
 and there were only limited data of $B_c$ available.
 Now the Large Hadron Collider (LHC)
will produce as much as $5\times10^{10}$ $B_c$ events per year~\cite{lhc1,lhc2}.
The huge amount of $B_c$ events will provide us a chance
to study $B_c$ decay to $3S,~4S$ states,
some channels also provide an opportunities to discover the new particles in $B_c$ decay.

The mesons can be described by the B-S equation.
Ref.~\cite{robert1} took the B-S equation to describe the light mesons $\pi$ and $K$,
then they calculated the mass and decay constant of $\pi$ by the B-S amplitudes~\cite{robert2},
they also studied the weak decays~\cite{robert3} and the strong decays~\cite{robert4} combine the Dyson-Schwinger
equation.
But in this paper, we describe the properties of heavy mesons and
the matrix elements of weak currents by improved B-S method,
which include two improvement~\cite{BS1}: one is about relativistic wavefunctions
which describe bound states with definite quantum number,
and a relativistic form of wavefunctions are solutions of the full Salpeter equations.
The other one is about the matrix elements of weak-current which obtained with
relativistic wavefunctions as input.
So the improved B-S method is good to describe the properties and decays of the heavy mesons with the relativistic corrections.

The paper is organized as follows.
In Sec.~II, we give the formulations of the exclusive semileptonic and
nonleptonic decays;
We show the hadronic weak-current matrix elements which is related to the wavefunctions of initial mesons and
final mesons in Section.~III;
We show the wavefunctions of initial and final mesons in Sec.~IV;
The corresponding results and conclusions are present in Sec.~V;
Finally in Appendix, we introduce the instantaneous Bethe-Salpeter equation.

\section{The formulations of semileptonic decays and nonleptonic decay of $B_c$}
 In this section we present the formulations of
  semi-leptonic decay and nonleptonic decay of $B_c$ mesons to $X(3940)$ and $X(4160)$
which
are considered as $\eta_c(3S)$ and $\eta_c(4S)$ states.

\subsection{Semileptonic decay of $B_c$}
\begin{figure}
\centering
\includegraphics[height=5cm]{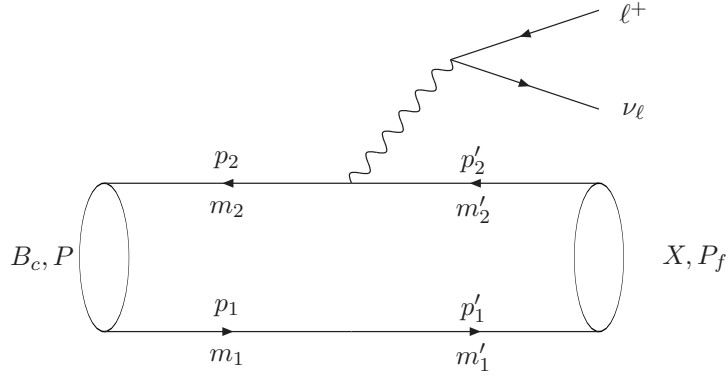}
\caption{\label{semileptonic}{Feynman diagram of the semi-leptonic decay $B_c\to X \ell^+\nu_\ell$,
$X$ denote $X(3940)$ or $X(4160)$.}}
\end{figure}

The feynman diagram of $B_c$ semileptonic decay to $X=X(3940)$ or $X=X(4160)$
is shown in Fig. \ref{semileptonic}.
The corresponding amplitude for the decay can be written as
\begin{eqnarray}\label{T}
T=\frac{G_F}{\sqrt{2}}V_{bc}\bar{u}_{\nu_\ell}\gamma^{\mu}(1-\gamma_5)
v_{\ell}\langle X(P_f)|J_{\mu}|B_c(P)\rangle\,,
\end{eqnarray}
where $V_{bc}$ is the CKM matrix element, $G_F$ is the the Fermi constant,
$J_{\mu}=V_{\mu}-A_{\mu}$ is the charged weak current, $P$ and
$P_f$ are the momentum of the initial meson $B_c$ and the final
state, respectively. The hadronic part can be written
as,
\begin{eqnarray}\label{form}
&&\langle
X(P_f)|V_{\mu}|B_c(P)\rangle=f_+(P+P_f)_{\mu}+f_-(P-P_f)_{\mu},
\nonumber\\
 &&\langle X(P_f)|A_{\mu}|B_c(P)\rangle=0,
\end{eqnarray}
where $f_+, f_-$ are the Lorentz invariant form factors.

We define $x\equiv E_\ell/M,\;\; y\equiv (P-P_f)^2/M^2$, where $E_\ell$
is the energy of the final charge lepton, $M$ is the mass of
initial meson. The differential width of the decay can be reduced
to:
\begin{eqnarray}\label{differ}
&&\frac{d^2\Gamma}{dxdy}=|V_{bc}|^2\frac{G_F^2M^5}{64{\pi}^3}
\nonumber\\
&&\left\{{\beta}_{++}\left[4\left(2x(1-\frac{M_f^2}{M^2}+y)-4x^2-y\right)
+\frac{m_\ell^2}{M^2}\left(8x+4\frac{M_f^2}{M^2}-3y-\frac{m_\ell^2}{M^2}\right)\right]\right.\nonumber\\
&&\left.({\beta}_{+-}+{\beta}_{-+})\frac{m_\ell^2}{M^2}
\left(2-4x+y-2\frac{M_f^2}{M^2}+\frac{m_\ell^2}{M^2}\right)
+{\beta}_{--}\frac{m_\ell^2}{M^2}\left(y-\frac{m_\ell^2}{M^2}\right)\right\}\,,
\end{eqnarray}
where $M_f$, $m_\ell$ are the masses of the meson and the lepton in
final states, respectively. $\beta_{++}=f^2_+$,
$\beta_{+-}=\beta_{-+}=f_+f_-$, $\beta_{--}=f^2_-$.

\subsection{Nonleptonic decay of $B_c$}

\begin{figure}
\centering
\includegraphics[height=5cm]{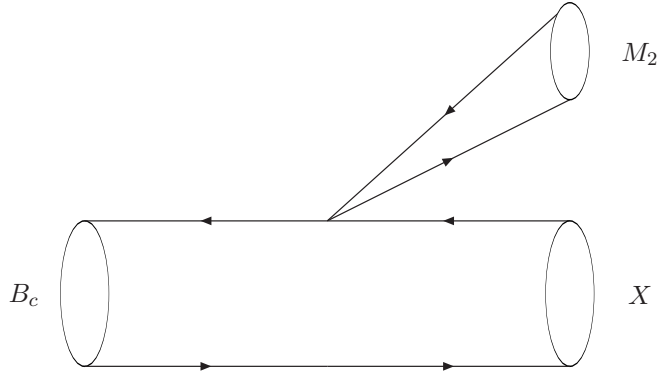}
\caption{\label{nonleptonic}{Feynman diagram of the nonleptonic decay $B_c\to X M_2$,
$X$ denote $X(3940)$ or $X(4160)$, $M_2$} denote a light meson: $\pi, K, \rho,$ or $K^*$.}
\end{figure}

For the nonleptonic decay of $B_c\to X+M_2$ in Fig.~\ref{nonleptonic},
the relevant effective Hamiltonian $H_{eff}$ is~\cite{Heff1,Heff2}:

\begin{eqnarray}\label{Heff}
H_{eff}=\frac{G_F}{\sqrt2}
\left\{V_{bc}[c_1(\mu)O_1^{bc}+c_2(\mu)O_2^{bc}]+h.c.\right\},
\end{eqnarray}
where $G_F$ is the Fermi constant, $V_{bc}$ is the CKM matrix element and
$c_i(\mu)$ are the scale-dependent Wilson coefficients.
$O_i$ are the operators responsible for the decays constructed by four quark fields and
have the structure as follows:
\begin{eqnarray}\label{O1O2}
&&O_1^{bc}=[V_{ud}(\bar d_\alpha u_\alpha)_{V-A}+V_{us}(\bar s_\alpha u_\alpha)_{V-A}](\bar c_\beta b_\beta)_{V-A},
\nonumber\\
&&O_2^{bc}=[V_{ud}(\bar d_\alpha u_\beta)_{V-A}+V_{us}(\bar s_\alpha u_\beta)_{V-A}](\bar c_\beta b_\alpha)_{V-A},
\end{eqnarray}
where $(\bar q_1q_2)_{V-A}=\bar q_1\gamma^\mu(1-\gamma_5)q_2$.

Since this is the primary study of these nonleptonic decays,
we apply the naive factorization to $H_{eff}$ \cite{naive},
the nonleptonic two-body decay amplitude $T$ can be reduce to a product of
a transition matrix element of a weak current $\langle X|J^\mu|B_c\rangle$
and an annihilation matrix element of another weak current $\langle M_2|J_\mu|0\rangle$:
\begin{eqnarray}\label{nonamplitude}
T=\langle XM_2|H_{eff}|B_c\rangle\approx
\frac{G_F}{\sqrt2}V_{bc}V_{ij}a_1\langle X|J^\mu|B_c\rangle\langle M_2|J_\mu|0\rangle,
\end{eqnarray}
$a_1=c_1+\frac{1}{N_c}c_2$ and $N_c=3$ is the number of colors.
While the annihilation matrix element $\langle M_2|J_\mu|0\rangle$ is related to
the decay constant of $M_2$. When $M_2$ is a pseudoscalar meson~\cite{pseudo},
$$\langle M_2|J_\mu|0\rangle=if_{M_2}P_{M_2\mu}$$
$f_{M_2}$ is the decay constant of meson $M_2$, $P_{M_2}$ is the momentum of $M_2$.
When $M_2$ is a vector meson~\cite{vector},
$$\langle M_2|J_\mu|0\rangle=\epsilon_\mu f_{M_2}M_{M_2}$$
where $M_{M_2}$, $f_{M_2}$ and $\epsilon$ are the mass, decay constant and polarization vector of
the vector meson $M_2$, respectively. The decay constant of the meson can be obtained
either by theoretical model or by indirect experiment measurement.

In Eq.~(\ref{differ}) and Eq.~(\ref{nonamplitude}),
we find that the most important things to get the decay width of the corresponding decay
are to calculate hadronic weak-current matrix elements $\langle
X(P_f)|J_{\mu}|B_c(P)\rangle$.
We will give the detailed calculation of the hadronic weak-current matrix elements in the Section.~III.

\section{The hadronic weak-current matrix elements}

The calculation of the hadronic weak-current matrix element are different from
model to model. In this paper, we combine the B-S method which is
based on relativistic B-S equation with Mandelstam formalism~\cite{Mand}
and relativistic wave functions to calculate the
hadronic matrix element.
The numerical values of wavefunctions
have been obtained by solving the full Salpeter equation which we
will introduce in Appendix. As an example, we consider the
semileptonic decay $B_c\to X\ell^+{\nu_\ell}$ in Fig.~\ref{semileptonic}. In this way,
at the leading order the hadronic matrix element can be written as
an overlapping integral over the wavefunctions of initial and final mesons~\cite{BS1},
\begin{eqnarray}\label{matrix}
\langle X(P_f)|J_{\mu}|B_c(P)\rangle=
\int\frac{d{\vec{q}}}{(2\pi)^3}{\rm Tr}\left[
\bar{\varphi}^{++}_{_{P_{f}}}(\vec {q}_{_f})\frac{\not\!P}{M}
{\varphi}^{++}_{_P}({\vec{q}})\gamma_{\mu}(1-\gamma_5)\right]\, ,
\end{eqnarray}
where $\vec{q}$ ($\vec{q}_{_f}$) is the relative three-momentum
between the quark and anti-quark in the initial (final) meson and
$\vec{q}_{_f}=\vec{q}-\frac{m'_1}{m'_1+m'_2}{\vec{P_f}}$.
$M$ is the mass of $B_c$, ${\vec{P_f}}$ is the three dimensional
momentum of $X$, ${\varphi}^{++}_P(\vec q)$ is the positive
Salpeter wavefunction of $B_c$ meson and
${\varphi}^{++}_{P_f}(\vec q_f)$
is the positive Salpeter wavefunction of $X$ meson,
$\bar{\varphi}^{++}_{_{P_f}}=\gamma_0({\varphi}^{++}_{_{P_f}})^{\dagger}\gamma_0$.
We will show the Salpeter wavefunctions for the different mesons in next section.

\section{The Relativistic Wavefunctions of Pseudoscalar Meson}

\subsection{ For $B_c$ meson with quantum numbers
$J^{P}=0^{-}$}

The general form for the relativistic wavefunction of
pseudoscalar meson $B_c$ can be written as~\cite{w1}:
\begin{eqnarray}\label{aa01}
\varphi_{0^-}(\vec q)&=&\Big[f_1(\vec q){\not\!P}+f_2(\vec q)M+
f_3(\vec q)\not\!{q_\bot}+f_4(\vec q)\frac{{\not\!P}\not\!{q_\bot}}{M}\Big]\gamma_5,
\end{eqnarray}
where $M$ is the mass of the pseudoscalar meson, and
$f_i(\vec q)$ are functions of $|\vec q|^2$. Due to
the last two equations of Eq.~(\ref{eq11}):
$\varphi_{0^-}^{+-}=\varphi_{0^-}^{-+}=0$, we have:
\begin{eqnarray}\label{constrain}
f_3(\vec q)&=&\frac{f_2(\vec q)
M(-\omega_1+\omega_2)}{m_2\omega_1+m_1\omega_2},~~~
f_4(\vec q)=-\frac{f_1(\vec q)
M(\omega_1+\omega_2)}{m_2\omega_1+m_1\omega_2}.
\end{eqnarray}
where $m_1, m_2$ and
$\omega_1=\sqrt{m_1^{2}+\vec{q}^2},\omega_2=\sqrt{m_2^{2}+\vec{q}^2}$ are
the masses and the energies of
 quark and anti-quark in $B_c$ mesons, $q_{_\bot}=q-(q\cdot P/M^2)P$, and $q_{\bot}^2=-|\vec q|^2$.

\begin{figure}[htbp]
\centering
\includegraphics[height=5cm]{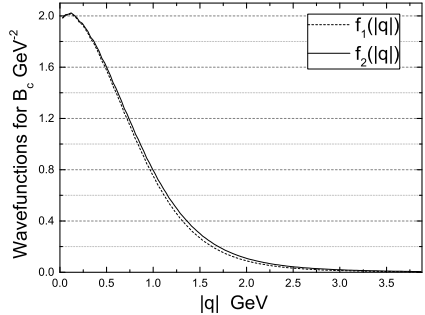}
\caption{\label{Bcfunction}The wavefunctions of $B_c$.}
\end{figure}

The numerical values of radial wavefunctions $f_1$, $f_2$ and
eigenvalue $M$ can be obtained by solving the first two Salpeter equations in
 Eq.~(\ref{eq11}). To show the numerical
results of wavefunctions explicitly, we plot the wavefunctions
of $B_c$ meson in
Fig.~\ref{Bcfunction}.
According to the Eq.~(\ref{eq10}) the relativistic positive wavefunction
of pseudoscalar meson $B_c$ in C.M.S can be written as \cite{w1}:
\begin{eqnarray}\label{0-postive}
{\varphi}^{++}_{0^-}(\vec{q})=b_1
\left[b_2+\frac{\not\!{P}}{M}+b_3\not\!{q_{\bot}}
+b_4\frac{\not\!{q_{\bot}}\not\!{P}}{M}\right]{\gamma}_5,
\end{eqnarray}
where the $b_i$s ($i=1,~2,~3,~4$) are related to the original
radial wavefunctions $f_1$, $f_2$, quark masses $m_1$, $m_2$, quark energy $w_1$, $w_2$,
and meson mass $M$:
$$b_1=\frac{M}{2}\left({f}_{1}(\vec{q})
+{f}_{2}(\vec{q})\frac{m_1+m_2}{\omega_1+\omega_2}\right),
b_2=\frac{\omega_1+\omega_2}{m_1+m_2}, b_3=-\frac{(m_1-m_2)}{m_1\omega_2+m_2\omega_1},
b_4=\frac{(\omega_1+\omega_2)}{(m_1\omega_2+m_2\omega_1)}.$$

\subsection{For $X(3940)$ and $X(4160)$ mesons with quantum numbers
$J^{P}=0^{-}$}

Because the $X(3940)$ and $X(4160)$ mesons have the same quantum numbers as
$B_c$, the wavefunctions of $X(3940)$ and $X(4160)$ mesons are similar to
Eq.~(\ref{0-postive}),
\begin{eqnarray}\label{final-2S}
{\varphi}^{++}_{_{P_f}}(\vec{q}_{f})=a_1
\left[a_2+\frac{\not\!{P_f}}{M_f}+a_3\not\!{q_{_{f\bot}}}
+a_4\frac{\not\!{q_{_{f\bot}}}\not\!{P_f}}{M_f}\right]{\gamma}_5,
\end{eqnarray}
$$a_1=\frac{M_f}{2}\left(f'_1(\vec{q}_f)
+f'_2(\vec{q}_f)\frac{m'_1+m'_2}{\omega'_1+\omega'_2}\right),
a_2=\frac{\omega'_1+\omega'_2}{m'_1+m'_2}, a_3=\frac{-m'_1+m'_2}{m'_1\omega'_2+m'_2\omega'_1},
a_4=\frac{\omega'_1+\omega'_2}{m'_1\omega'_2+m'_2\omega'_1}.$$

\begin{figure}[htbp]
\centering
\includegraphics[height=5cm]{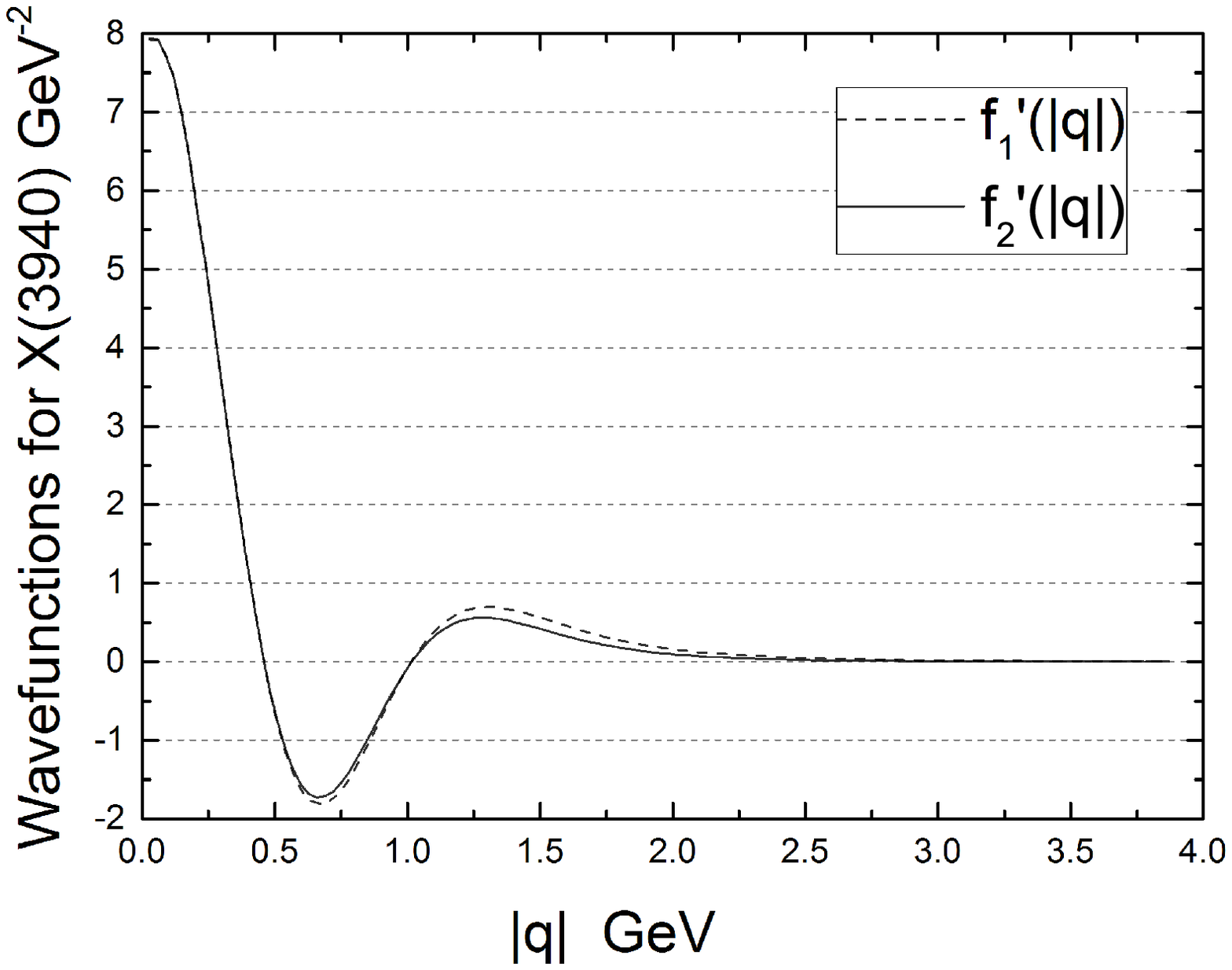}
\includegraphics[height=5cm]{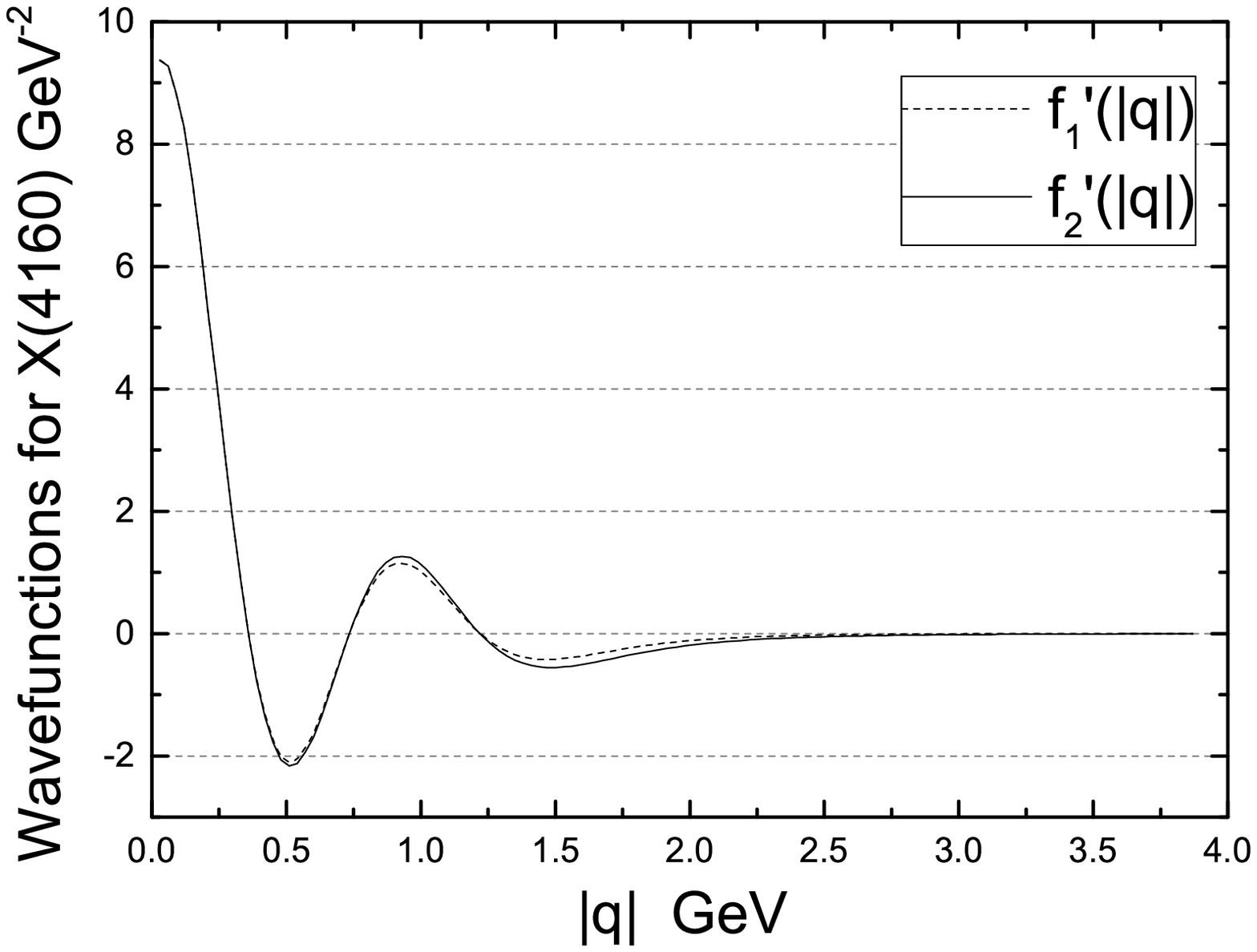}
\caption{\label{wavefunction}The wavefunctions of $X(3940)$ and $X(4160)$.}
\end{figure}

Where $M_f$, $P_f$, $f'_i(\vec q_f)$ are the mass,
momentum and the radial wavefunctions of $X(3940)$ and $X(4160)$, respectively.
$m'_1, m'_2$ and
$\omega'_1=\sqrt{m_1^{\prime2}+\vec{q}_f^2},\omega'_2=\sqrt{m_2^{\prime2}+\vec{q}_f^2}$ are
the masses and the energies of
 quark and anti-quark in $X(3940)$ and $X(4160)$.
To show the numerical
results of wavefunctions explicitly, we plot the wavefunctions
of $X(3940)$ and $X(4160)$ states in
Fig.~\ref{wavefunction}.

\section{Number results and discussions}
\subsection{Semi-leptonic decays}
In order to fix Cornell potential in Eq.(\ref{eq16}) and masses of quarks,
 we take these parameters: $a=e=2.7183,
\lambda=0.21$ GeV$^2$, ${\Lambda}_{QCD}=0.27$ GeV, $\alpha=0.06$
GeV, $m_b=4.96$ GeV, $m_c=1.62$ GeV, $etc$~\cite{mass1},
 which are best to fit the mass spectra of ground states $B_c$ and other heavy mesons.
Taking these parameters to B-S equation,
and solving the B-S equation numerically,
we get the masses of $X(3940)$, $X(4160)$ and $B_c$ as:
$M_{X(3940)}$=3.942 GeV, $M_{X(4160)}$=4.156 GeV, $M_{B_c}=6.276$ GeV.

\begin{figure}[htbp]
\centering
\includegraphics[height=5cm]{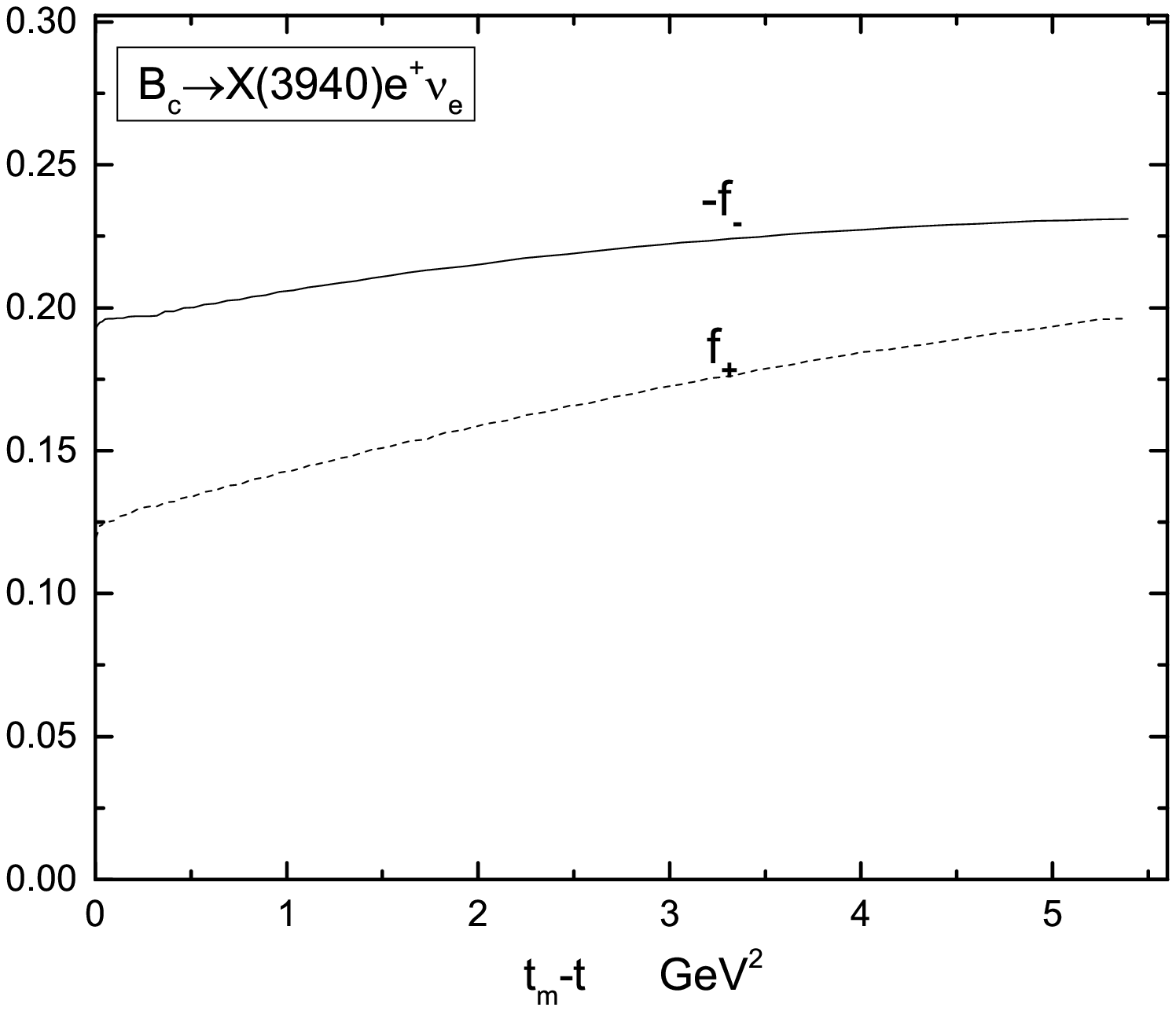}
\includegraphics[height=5cm]{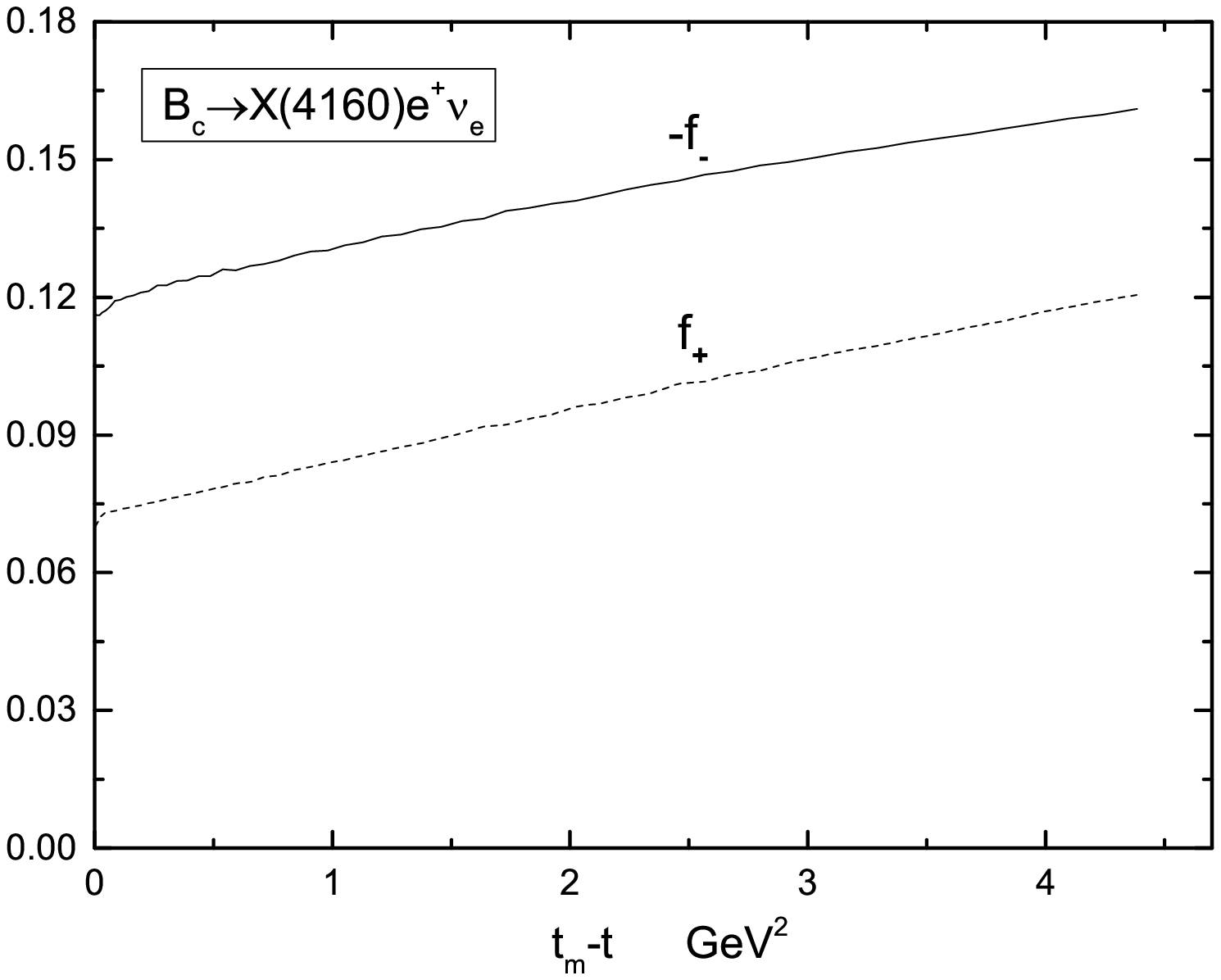}
\caption{\label{formfactor}The form factor of semileptonic decay $B_c$ to $X(3940)$ and $X(4160)$.}
\end{figure}
\begin{figure}[htbp]
\centering
\includegraphics[height=5cm]{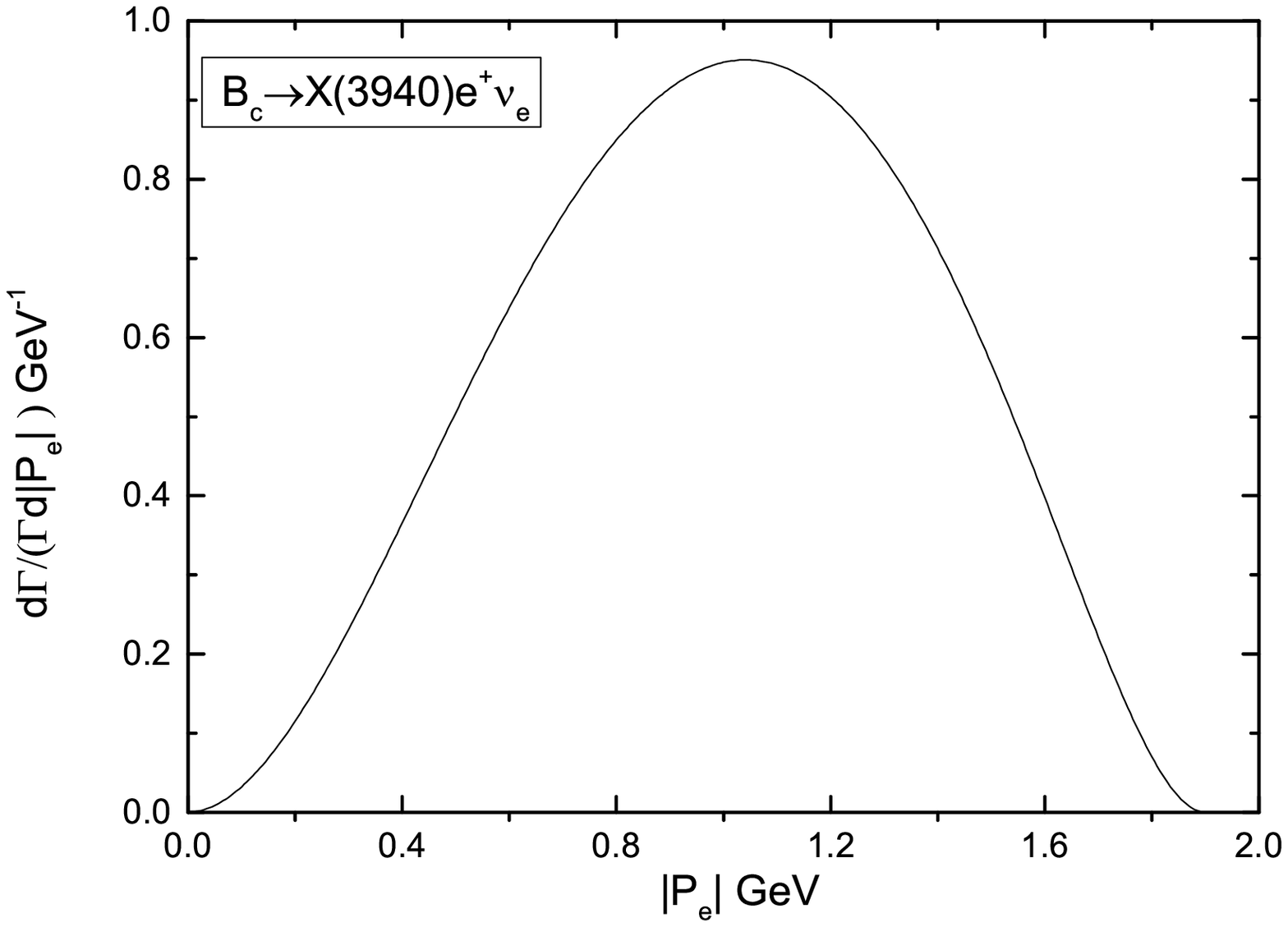}
\includegraphics[height=5cm]{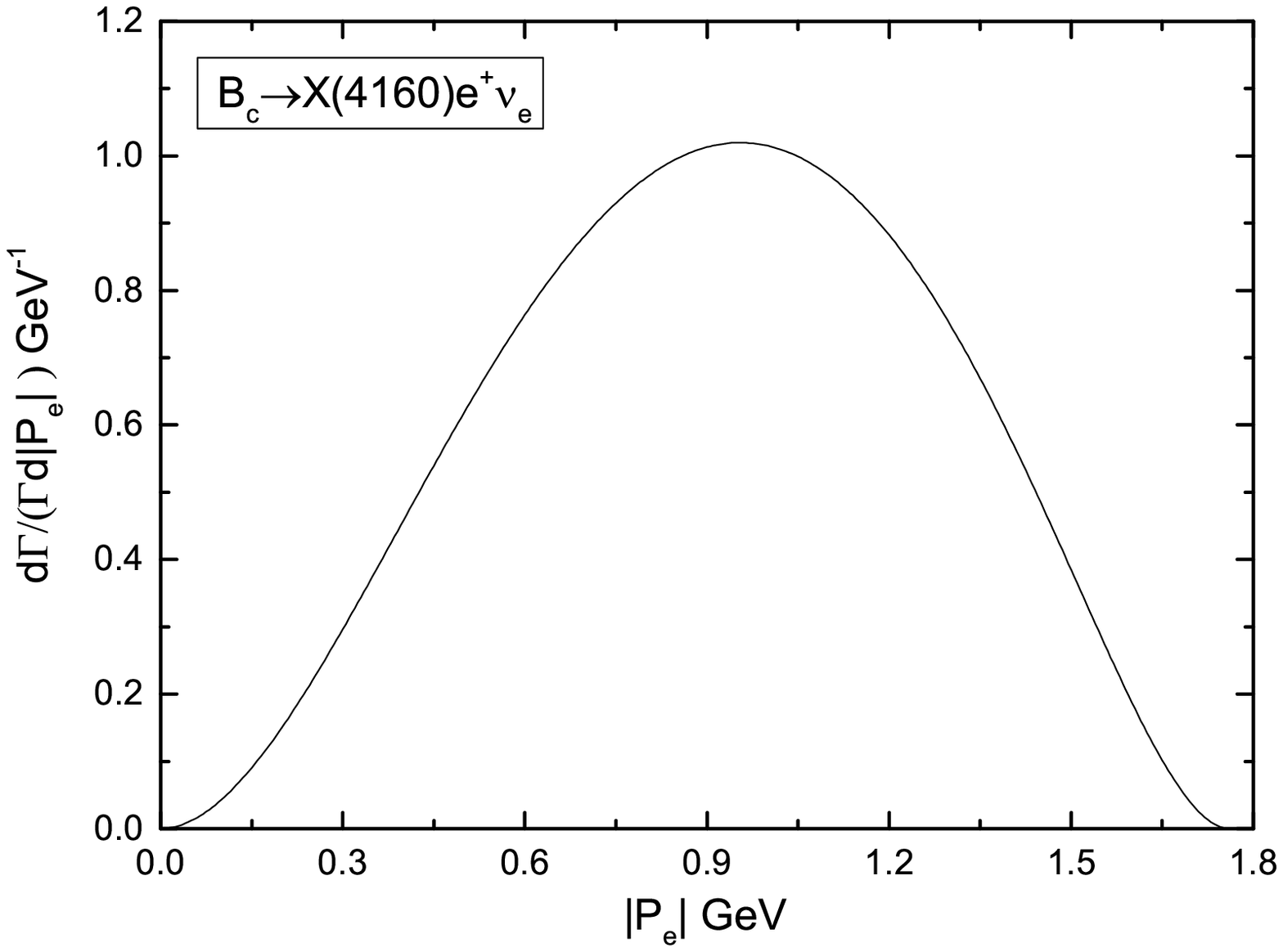}
\caption{\label{energyspectra}The leptonic energy spectra of semileptonic decay $B_c$ to $X(3940)$ and $X(4160)$.}
\end{figure}

For semileptonic decays we need to input
the CKM matrix elements:
 $V_{cb}=0.0406$,
$G_F=1.166\times10^{-5}$ GeV$^{-2}$
and the life time of $B_c$ meson: $\tau_{B_c} =
0.453$ps,
which are taken from PDG~\cite{PDG}.
In Section.~III,
we have found that the hadronic weak-current matrix element is overlapping integral over the
wavefunctions of initial and final states,
and the hadronic weak-current matrix element can be written as the form factors $f_+$ and $f_-$.
The form factors are relate to four-momentum transfer squared $t=(P-P_f)^2=M^2+M_f^2-2ME_f$
which provides the kinematic range for the semileptonic decay of $B_c$.
It varies from $t=0$ to $t=5.45$ GeV$^2$ for the decays to $X(3940)$ and
from $t=0$ to $t=4.48$ GeV$^2$ for the decays to $X(4160)$.
We give the relations of $(t_m-t)$($t_m=(M-M_f)^2$ is the maximum of $t$) and the form factors,
which are calculated by Eq.~(\ref{matrix}) in Fig.~\ref{formfactor}.
The leptonic energy spectra $\frac{d\Gamma}{\Gamma dP_e}$ for semileptonic $B_c$
decay to $X(3940)$ and $X(4160)$ are calculated by Eq.~(\ref{differ}).
The results are plot in Fig.~\ref{energyspectra}.
In Table.~\ref{semidecaywidth}, we summarize the decay widths of the semileptonic
$B_c \to X\ell^+ \nu_\ell$ ($X=X(3940)$ or $X(4160)$, $\ell=e,\mu,\tau$).
We have taken $\Gamma_e\simeq\Gamma_\mu$ with the massless lepton limit since
the muon mass effect is negligible for these transitions with large kinematic ranges.
The semileptonic decay widths of $B_c^+\to X(3940)$ are larger than $B_c^+\to X(4160)$ in Table.~\ref{semidecaywidth},
there are two reasons: first, the former decay has larger kinematic ranges,
second, there is one minus part in the wavefunctions of $X(3940)$,
and there are two minus parts in the wavefunctions of $X(4160)$ in Fig.~\ref{wavefunction},
after the overlapping integral in Eq.~(\ref{matrix}),
much more minus parts of the wavefunctions cause the smaller result for $X(4160)$.

\begin{table}[htbp]
\caption{\label{semidecaywidth}The decay widths of exclusive semileptonic decay $B_c$ to $X(3940),~X(4160)$ (in $10^{-15}$GeV).}
\begin{center}
\begin{tabular}{|c|c|c|c|}
\hline \hline
Mode&Ours&Mode&Ours  \\
\hline
 $B_c^+\to$ $X(3940)$$e^+\nu_e$&0.147&$B_c^+\to$$X(4160)$$e^+\nu_e$&3.46$\times10^{-2}$\\
$B_c^+\to$ $X(3940)$$\tau^+\nu_\tau$&$4.35\times 10^{-3}$&$B_c^+\to$$X(4160)$$\tau^+\nu_\tau$&$2.57\times10^{-4}$\\
\hline \hline
\end{tabular}
\end{center}
\end{table}

For the exclusive nonleptonic decay, we only consider two body decays,
and another meson is light meson. The corresponding CKM matrix elements are:
$V_{ud}=0.974$ and $V_{us}=0.2252$. The masses and decay constants are:
$M_\pi=0.140$ GeV,      $f_\pi=0.130$ GeV,
$M_\rho=0.775$ GeV,      $f_\rho=0.205$ GeV,
$M_K=0.494$ GeV,      $f_K=0.156$ GeV,
$M_{K^*}=0.892$ GeV,      $f_{K^*}=0.217$ GeV~\cite{PDG,kk}, respectively.
The kinematic range of nonleptonic decay is fixed value,
so the form factors of $B_c$ nonleptonic decay are definite value.
Using the form factors of $B_c$ nonleptonic decay and the decay constants,
we show the nonleptonic decay widths which are related to the parameter $a_1$ in Table.~\ref{nondecaywidth}.
The results of $B_c$ nonleptonic decay are affected by the CKM matrix elements, so the results of light mesons $\pi, \rho$
are larger than the ones of light mesons $K, K^*$ in Table.~\ref{nondecaywidth}, respectively.

\begin{table}[htbp]
\caption{\label{nondecaywidth}The decay widths of exclusive nonleptonic decay $B_c$ to $X(3940),~X(4160)$ (in $10^{-15}$GeV).}
\begin{center}
\begin{tabular}{|c|c|c|c|}
\hline \hline
Mode&Ours&Mode&Ours  \\
\hline
 $B_c^+\to$ $X(3940)$+$\pi$&7.57$\times 10^{-2}a^2_1$&$B_c^+\to$$X(4160)$+$\pi$&2.29$\times 10^{-2}a^2_1$\\
$B_c^+\to$$X(3940)$+$K$&5.51$\times 10^{-3}a^2_1$&$B_c^+\to$$X(4160)$+$K$&1.61$\times 10^{-3}a^2_1$\\
$B_c^+\to$$X(3940)$+$\rho$&0.149$a^2_1$&$B_c^+\to$$X(4160)$+$\rho$&4.17$\times 10^{-2}a^2_1$\\
$B_c^+\to$$X(3940)$+$K^*$&8.30$\times 10^{-3}a^2_1$&$B_c^+\to$$X(4160)$+$K^*$&2.22$\times 10^{-3}a^2_1$\\
\hline \hline
\end{tabular}
\end{center}
\end{table}

\begin{figure}[htbp]
\centering
\includegraphics[height=5cm]{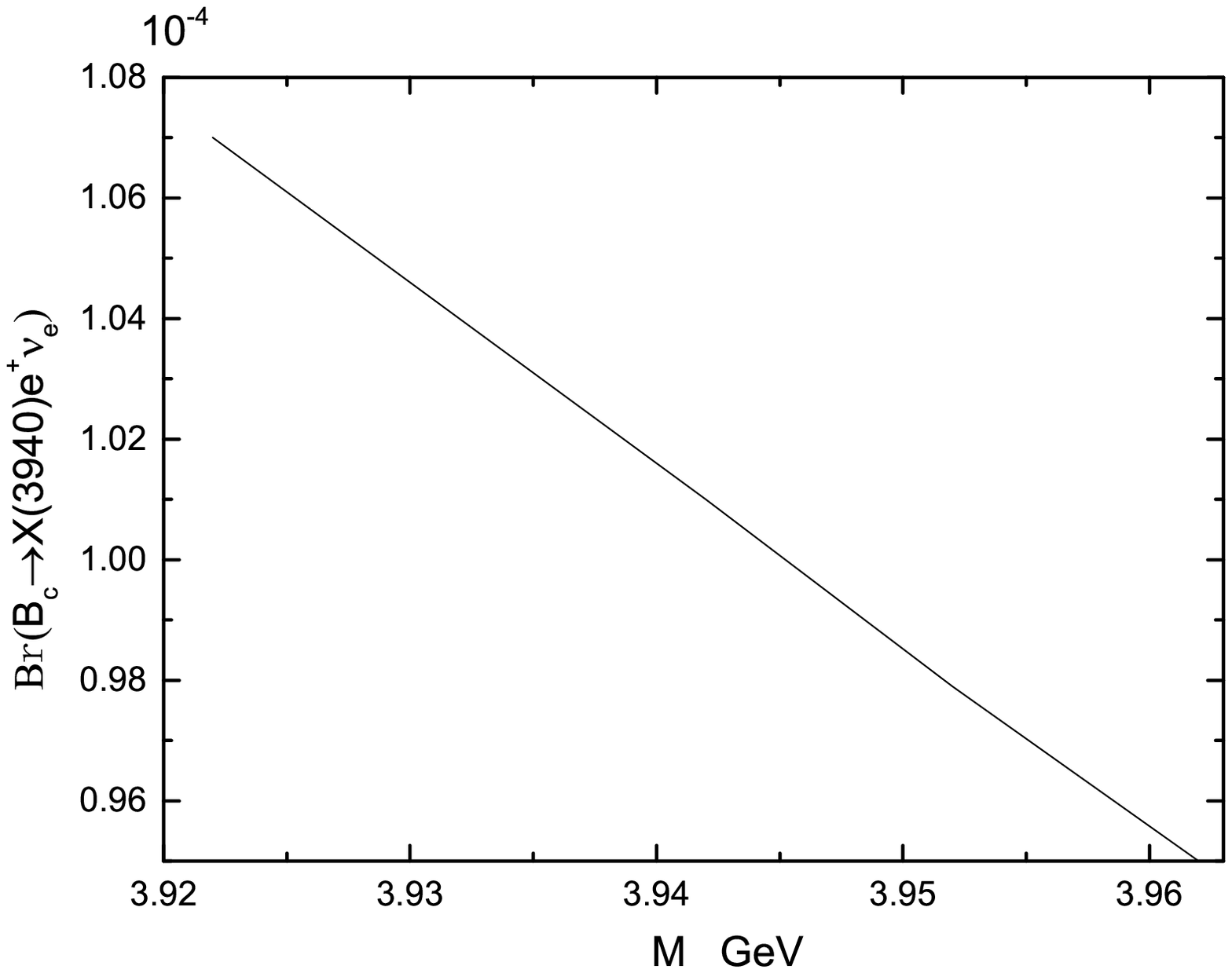}
\includegraphics[height=5cm]{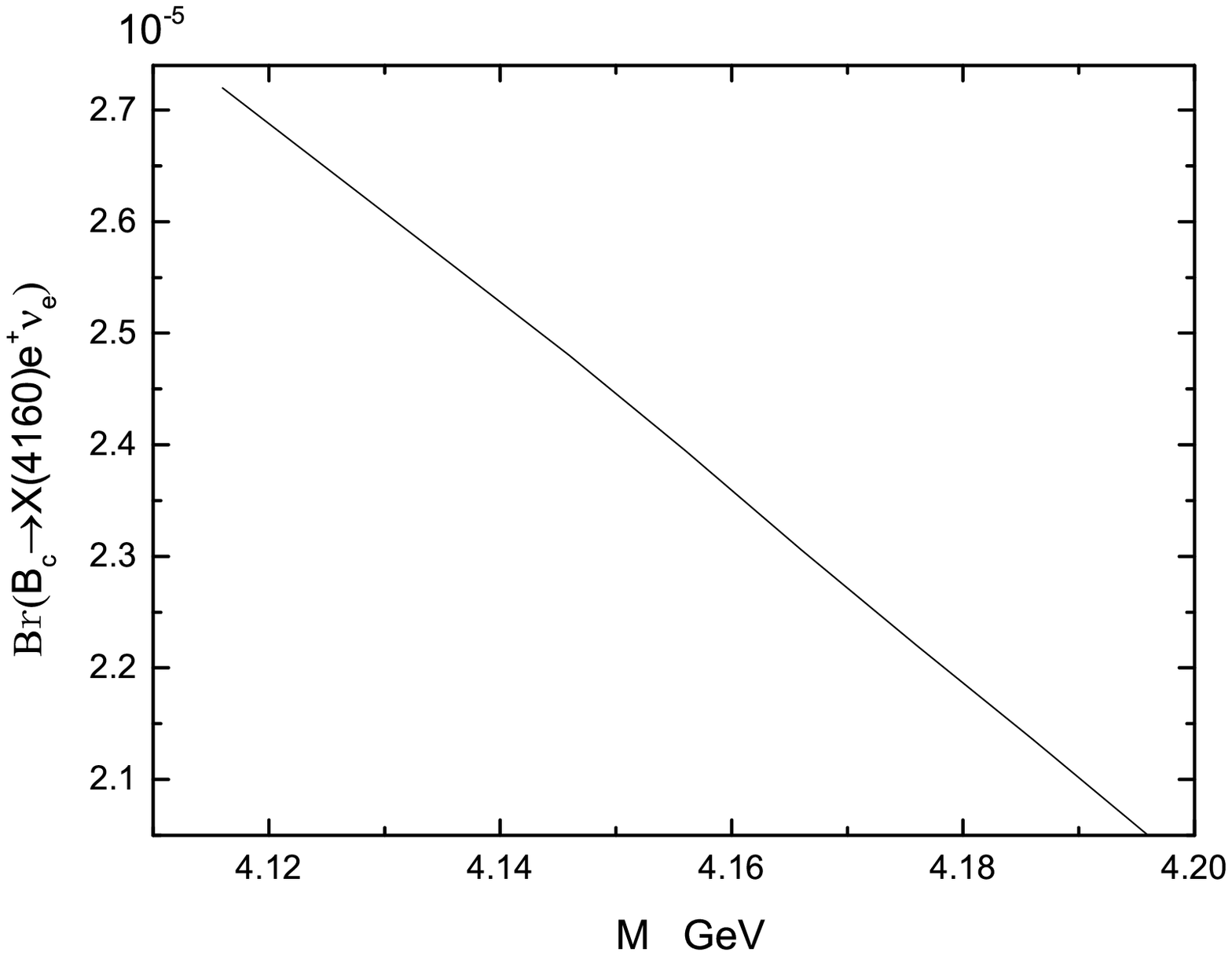}
\caption{\label{M-Br}The relations of branching ratios to the mass of final mesons.}
\end{figure}

In order to compare the numerical values of semileptonic and nonleptonic decays,
we show the branching ratios of semileptonic and nonleptonic $B_c$ with $a_1=1.14$~\cite{Heff1,Heff2} in Table.~\ref{Branch}.
We find that the central value results of $B_c^+\to$ $X(3940)$$e^+\bar\nu_e$ and $B_c^+\to X(3940)\tau^+\bar\nu_\tau$ are
less than Ref.~\cite{Lu}. But considering the errors of results, our results are in accordance with ones of Ref.~\cite{Lu}.
If we compare $B_c\to$$X(3940)$, $X(4160)$ in this paper with $B_c$ decays to $\eta_c(1S)$ in literatures,
for example in Ref.~\cite{bc1,bc2,bc13},
our results are almost two order smaller than the results of $B_c$ decay to $\eta_c(1S)$.
there are two reasons, one reason is that the $B_c\to$$X(3940)$, $X(4160)$ have small kinematic ranges,
another one is that the wavefunctions have some minus parts in $X(3940)$, and $X(4160)$.
Because the mass errors of $X(3940)$ and $X(4160)$ are still large,
the widths and branching ratios of $B_c$ weak decays to $X(3940)$ and $X(4160)$ are influenced
by the masses of $X(3940)$ and $X(4160)$,
we plot the relations of the branching ratios of $B_c$ weak decays to $X(3940)$ and $X(4160)$ to
the masses of $X(3940)$ and $X(4160)$ in Fig.~\ref{M-Br}.
The relations of the branching ratios to the masses of $X(3940)$ and $X(4160)$ are linear.
The branching ratios decrease with the increase of the masses of $X(3940)$ and $X(4160)$.
\begin{table}[htbp]
\caption{\label{Branch}The branching ratio(in $\%$) of exclusive semileptonic and nonleptonic decay $B_c$ to $X(3940),~X(4160)$ with $a_1=1.14$.}
\begin{center}
\begin{tabular}{|c|c|c|c|c|}
\hline \hline
Mode&Ours&\cite{Lu}&Mode&Ours \\
\hline
 $B_c^+\to$ $X(3940)$$e^+\nu_e$&$1.02\times 10^{-2}$&$1.9^{+0.2+0.1+0.0+0.8+0.7+0.0}_{-0.1-0.1-0.0-0.9-0.7-0.0}\times10^{-2}$
 &$B_c^+\to$$X(4160)$$e^+\nu_e$&2.39$\times 10^{-3}$\\
$B_c^+\to$$X(3940)$$\tau^+\nu_\tau$&3.00$\times 10^{-4}$&$5.7^{+0.6+0.7+0.3+2.4+2.0+0.0}_{-0.3-0.4-0.3-2.7-2.2-0.1}\times10^{-4}$
&$B_c^+\to$$X(4160)$$\tau^+\nu_\tau$&1.78$\times 10^{-5}$\\
 $B_c^+\to$$X(3940)$+$\pi$&6.78$\times 10^{-3}$&--&$B_c^+\to$$X(4160)$+$\pi$&2.05$\times 10^{-3}$\\
$B_c^+\to$$X(3940)$+$K$&4.94$\times 10^{-4}$&--&$B_c^+\to$$X(4160)$+$K$&1.44$\times 10^{-4}$\\
$B_c^+\to$$X(3940)$+$\rho$&1.34$\times 10^{-2}$&--&$B_c^+\to$$X(4160)$+$\rho$&3.73$\times 10^{-3}$\\
$B_c^+\to$$X(3940)$+$K^*$&7.44$\times 10^{-4}$&--&$B_c^+\to$$X(4160)$+$K^*$&2.00$\times 10^{-4}$\\
\hline \hline
\end{tabular}
\end{center}
\end{table}

In conclusion, considering $X(3940)$ and $X(4160)$ as $\eta_c(3S)$ and $\eta_c(4S)$ states,
we study the semileptonic and nonleptonic $B_c$ decays to $X(3940)$ and $X(4160)$
by the improved B-S method.
The corresponding decay form factors are calculated and the corresponding
decay widths and branching ratios are obtained.
The exclusive decay widths and branching ratios are very small,
because of the minus value in the wavefunctions.
But now the Large Hadron Collider (LHC)
will produce as much as $5\times10^{10}$ $B_c$ events per year~\cite{lhc1,lhc2},
if we can observe the sufficient events, some channels will provide us a sizable ratios,
and may be we will detect the productions of $X(3940)$ and $X(4160)$ in $B_c$ exclusive
weak semileptonic and nonleptonic decay.
That will provide us a new way to observe the $X(3940)$ and $X(4160)$ in the future.

 \noindent
{\Large \bf Acknowledgements}
This work was supported in part by
the National Natural Science Foundation of China (NSFC) under
Grant No.~11405004, No.~11405037, No.~11505039 and No.~11575048.

\appendix{
\section{Instantaneous Bethe-Salpeter Equation}

In this section, we briefly review the Bethe-Salpeter equation and
its instantaneous one, the Salpeter equation.

The BS equation is read as~\cite{BS}:
\begin{equation}
(\not\!{p_{1}}-m_{1})\chi(q)(\not\!{p_{2}}+m_{2})=
i\int\frac{d^{4}k}{(2\pi)^{4}}V(P,k,q)\chi(k)\;, \label{eq1}
\end{equation}
where $\chi(q)$ is the BS wave function, $V(P,k,q)$ is the
interaction kernel between the quark and antiquark, and $p_{1},
p_{2}$ are the momentum of the quark 1 and anti-quark 2.

We divide the relative momentum $q$ into two parts,
$q_{\parallel}$ and $q_{\perp}$,
$$q^{\mu}=q^{\mu}_{\parallel}+q^{\mu}_{\perp}\;,$$
$$q^{\mu}_{\parallel}\equiv (P\cdot q/M^{2})P^{\mu}\;,\;\;\;
q^{\mu}_{\perp}\equiv q^{\mu}-q^{\mu}_{\parallel}\;.$$

B-S equation Eq.~(\ref{eq1}) is a four dimension covariant equation,
in order to solve the Eq.~(\ref{eq1}),
we will take the instantaneous approximation in the interaction kernel $V(P,k,q)$,
then the B-S equation will lose the covariance.
The effect of instantaneous approximation in $V(P,k,q)$ could be corrected by the
retardation effects in $V(P,k,q)$.
But the retardation effects in $V(P,k,q)$ are very small for the heavy mesons~\cite{eff1,eff2,eff3},
this means that the influence of the instantaneous approximation on the covariance of B-S equation
are very small for the heavy mesons.
The instantaneous approximation in $V(P,k,q)$ almost
don't influence the wavefunctions,
and the decay matrix elements which involve the heavy mesons mostly unchanged~\cite{eff1}.
Our model mostly keeps the covariance in the calculation, and the weak decay results also satisfy the Lorentz-covariance.

In instantaneous approach, the kernel $V(P,k,q)$ takes the simple
form~\cite{Salp}:
$$V(P,k,q) \Rightarrow V(|\vec k-\vec q|)\;.$$

Let us introduce the notations $\varphi_{p}(q^{\mu}_{\perp})$ and
$\eta(q^{\mu}_{\perp})$ for three dimensional wave function as
follows:
$$
\varphi_{p}(q^{\mu}_{\perp})\equiv i\int
\frac{dq_{p}}{2\pi}\chi(q^{\mu}_{\parallel},q^{\mu}_{\perp})\;,
$$
\begin{equation}
\eta(q^{\mu}_{\perp})\equiv\int\frac{dk_{\perp}}{(2\pi)^{3}}
V(k_{\perp},q_{\perp})\varphi_{p}(k^{\mu}_{\perp})\;. \label{eq5}
\end{equation}
Then the BS equation can be rewritten as:
\begin{equation}
\chi(q_{\parallel},q_{\perp})=S_{1}(p_{1})\eta(q_{\perp})S_{2}(p_{2})\;.
\label{eq6}
\end{equation}
The propagators of the two constituents can be decomposed as:
\begin{equation}
S_{i}(p_{i})=\frac{\Lambda^{+}_{ip}(q_{\perp})}{J(i)q_{p}
+\alpha_{i}M-\omega_{i}+i\epsilon}+
\frac{\Lambda^{-}_{ip}(q_{\perp})}{J(i)q_{p}+\alpha_{i}M+\omega_{i}-i\epsilon}\;,
\label{eq7}
\end{equation}
with
\begin{equation}
\omega_{i}=\sqrt{m_{i}^{2}+q^{2}_{_T}}\;,\;\;\;
\Lambda^{\pm}_{ip}(q_{\perp})= \frac{1}{2\omega_{ip}}\left[
\frac{\not\!{P}}{M}\omega_{i}\pm
J(i)(m_{i}+{\not\!q}_{\perp})\right]\;, \label{eq8}
\end{equation}
where $i=1, 2$ for quark and anti-quark, respectively,
 and
$J(i)=(-1)^{i+1}$.

Introducing the notations $\varphi^{\pm\pm}_{p}(q_{\perp})$ as:
\begin{equation}
\varphi^{\pm\pm}_{p}(q_{\perp})\equiv
\Lambda^{\pm}_{1p}(q_{\perp})
\frac{\not\!{P}}{M}\varphi_{p}(q_{\perp}) \frac{\not\!{P}}{M}
\Lambda^{{\pm}}_{2p}(q_{\perp})\;. \label{eq10}
\end{equation}

With contour integration over $q_{p}$ on both sides of
Eq.~(\ref{eq6}), we obtain:
$$
\varphi_{p}(q_{\perp})=\frac{
\Lambda^{+}_{1p}(q_{\perp})\eta_{p}(q_{\perp})\Lambda^{+}_{2p}(q_{\perp})}
{(M-\omega_{1}-\omega_{2})}- \frac{
\Lambda^{-}_{1p}(q_{\perp})\eta_{p}(q_{\perp})\Lambda^{-}_{2p}(q_{\perp})}
{(M+\omega_{1}+\omega_{2})}\;,
$$
and the full Salpeter equation:
$$
(M-\omega_{1}-\omega_{2})\varphi^{++}_{p}(q_{\perp})=
\Lambda^{+}_{1p}(q_{\perp})\eta_{p}(q_{\perp})\Lambda^{+}_{2p}(q_{\perp})\;,
$$
$$(M+\omega_{1}+\omega_{2})\varphi^{--}_{p}(q_{\perp})=-
\Lambda^{-}_{1p}(q_{\perp})\eta_{p}(q_{\perp})\Lambda^{-}_{2p}(q_{\perp})\;,$$
\begin{equation}
\varphi^{+-}_{p}(q_{\perp})=\varphi^{-+}_{p}(q_{\perp})=0\;.
\label{eq11}
\end{equation}

For the different $J^{PC}$ (or $J^{P}$) states, we give the general form of
wave functions. Reducing the wave functions by the last
equation of Eq.~(\ref{eq11}), then solving the first and second equations in Eq.~(\ref{eq11}) to
get the wave functions and mass spectrum. We have discussed the
solution of the Salpeter equation in detail in Ref.~\cite{w1,mass1}.

The normalization condition for BS wave function is:
\begin{equation}
\int\frac{q_{_T}^2dq_{_T}}{2{\pi}^2}Tr\left[\overline\varphi^{++}
\frac{{/}\!\!\!
{P}}{M}\varphi^{++}\frac{{/}\!\!\!{P}}{M}-\overline\varphi^{--}
\frac{{/}\!\!\! {P}}{M}\varphi^{--}\frac{{/}\!\!\!
{P}}{M}\right]=2P_{0}\;. \label{eq12}
\end{equation}

 In our model, the instantaneous interaction kernel $V$ is Cornell
potential, which is the sum of a linear scalar interaction and a vector interaction:
\begin{equation}\label{vrww}
V(r)=V_s(r)+V_0+\gamma_{_0}\otimes\gamma^0 V_v(r)= \lambda
r+V_0-\gamma_{_0}\otimes\gamma^0\frac{4}{3}\frac{\alpha_s}{r}~,
\end{equation}
 where $\lambda$ is the string constant and $\alpha_s(\vec
q)$ is the running coupling constant. In order to fit the data of
heavy quarkonia, a constant $V_0$ is often added to confine
potential. One can see that $V_v(r)$ diverges at $r=0$, we introduce a factor $e^{-\alpha r}$ to avoid
the divergence:
\begin{equation}
V_s(r)=\frac{\lambda}{\alpha}(1-e^{-\alpha r})~,
~~V_v(r)=-\frac{4}{3}\frac{\alpha_s}{r}e^{-\alpha r}~.
\end{equation}\label{vsvv}
 It is easy to
know that when $\alpha r\ll1$, the potential becomes to Eq.~(\ref{vrww}). In the momentum space and the C.M.S of the bound state,
the potential reads :
$$V(\vec q)=V_s(\vec q)
+\gamma_{_0}\otimes\gamma^0 V_v(\vec q)~,$$
\begin{equation}
V_s(\vec q)=-(\frac{\lambda}{\alpha}+V_0) \delta^3(\vec
q)+\frac{\lambda}{\pi^2} \frac{1}{{(\vec q}^2+{\alpha}^2)^2}~,
~~V_v(\vec q)=-\frac{2}{3{\pi}^2}\frac{\alpha_s( \vec q)}{{(\vec
q}^2+{\alpha}^2)}~,\label{eq16}
\end{equation}
where the running coupling constant $\alpha_s(\vec q)$ is :
$$\alpha_s(\vec q)=\frac{12\pi}{33-2N_f}\frac{1}
{\log (a+\frac{{\vec q}^2}{\Lambda^{2}_{QCD}})}~.$$ We introduce a small
parameter $a$ to
avoid the divergence in the denominator. The constants $\lambda$, $\alpha$, $V_0$ and
$\Lambda_{QCD}$ are the parameters that characterize the potential. $N_f=3$ for $\bar bq$ (and $\bar cq$) system.
}


\begin{thebibliography}{99}


\bibitem{FirstCharm}
K. Abe $et~ al.$, Belle Collaboration, $Phys.~Rev.~Lett.$ {\bf 89},
142001(2003).

\bibitem{3872}
S. K. Choi $et~ al.$, Belle Collaboration, $Phys.~Rev.~Lett.$ {\bf 91},
262001(2003).
\bibitem{3915}
S. Uehara $et~ al.$, Belle Collaboration, $Phys.~Rev.~Lett.$ {\bf 104}, 092001(2010).
\bibitem{3940}
K. Abe $et~ al.$, Belle Collaboration, $Phys.~Rev.~Lett.$ {\bf 98},
082001(2007).
\bibitem{39404160}
P. Pakhlov $et~ al.$, Belle Collaboration, $Phys.~Rev.~Lett.$ {\bf 100},
202001(2008).
\bibitem{PDG}
K. A. Olive $et~ al$., (Partile Data Group), $Chin.~Phys.~C.$ {\bf 38},
090001(2015).
\bibitem{liuxiang1}
X. Liu, $Chin.~Sci.~Bull.$ {\bf59}, 3815(2014).
\bibitem{zhushilin1}
H. X. Chen, W. Chen, X. Liu and S. L. Zhu, $Phys.~Rept.$, {\bf639} 1(2016).
\bibitem{stephen1}
S. L. Olsen, $Front.~Phys.$ {\bf10}, 101401(2015).
\bibitem{zhao1}
B. Q. Li and K. T. Chao, $Phys.~Rev.~D$ {\bf79}, 094004(2009).
\bibitem{kim}
H. C. Kim, K. S. Kim, M. K. Cheoun, D. Jido and M. Oka, arXiv:1602.07540 [hep-ph].
\bibitem{Lu}
Y. M. Wang and C. D. Lu, $Phys.~Rev.~D$ {\bf77}, 054003(2008).
\bibitem{39401}
Z. G. He and B. Q. Li, $Phys.~Lett.~B$ {\bf693}, 36--43(2010).
\bibitem{zhuruilin}
R. L. Zhu, $Phys.~Rev.~D$ {\bf92}, 074017(2015).
\bibitem{liuxiang2}
L. P. He, D. Y. Chen, X. Liu and T. Matsuki, $Eur.~Phys.~J.~C$ {\bf74}, 3028(2014).
\bibitem{41601}
Y. C. Yang, Z. R. Xia and J. L. Ping, $Phys.~Rev.~$D {\bf81}, 094003 (2010).
\bibitem{th6}
H. Wang, Z. Z. Yan and J. L. Ping, $Eur.~Phys.~J.~$C {\bf75}, 196 (2015).
\bibitem{bc1}
C. H. Chang, and Y. Q. Chen, $Phys.~Rev.~D$ {\bf49}, 3399(1994).
\bibitem{bc2}
C. H. Chang and Y. Q. Chen, $Phys.~Rev.~D$ {\bf46}, 3845 (1992).
\bibitem{bc3}
M. Beneke and G. Buchalla, $Phys.~Rev.~D$ {\bf53}, 4991 (1996).
\bibitem{bc4}
J. F. Liu and K. T. Chao, $Phys.~Rev.~D$ {\bf56}, 4133 (1997).
\bibitem{bc5}
Y. S. Dai and D. S. Du, $Eur.~Phys.~J.~C$ {\bf9}, 557 (1999).
\bibitem{Ebert11}
D. Ebert, R. N. Faustov and V. O. Galkin, $Phys.~Rev.~D$ {\bf82}, 034019 (2010).
\bibitem{Ebert12}
D. Ebert, R. N. Faustov and V. O. Galkin, $Phys.~Rev.~D$ {\bf82}, 034032 (2010).
\bibitem{Ebert13}
D. Ebert, R. N. Faustov and V. O. Galkin, $Phys.~Rev.~D$ {\bf68}, 094020 (2003).
\bibitem{Ebert14}
D. Ebert, R. N. Faustov and V. O. Galkin, $Eur.~Phys.~J.~C$ {\bf32}, 29 (2003).
\bibitem{Ebert15}
D. Ebert, R. N. Faustov and V. O. Galkin, $Phys.~Rev.~D$ {\bf67}, 014027 (2003).
\bibitem{Ivanov1}
M. A. Ivanov, J. G. K$\ddot{o}$rner and P. Santorelli, $Phys.~Rev.~D$, {\bf63}, 074010 (2001).
\bibitem{Ivanov2}
M. A. Ivanov, J. G. K$\ddot{o}$rner and P. Santorelli, $Phys.~Rev.~D$, {\bf73}, 054024 (2006).
\bibitem{Ivanov3}
M. A. Ivanov, J. G. K$\ddot{o}$rner and P. Santorelli, $Phys.~Rev.~D$, {\bf71}, 094006, (2005), {\bf75}, 019901, (2007).
\bibitem{Ivanov4}
M. A. Ivanov, J. G. K$\ddot{o}$rner and O. N. Pakhomova, $Phys.~Lett.~B$ {\bf555}, 189 (2003)
\bibitem{bc6}
 X. X. Wang, W. Wang and C. D. Lu, $Phys.~Rev.~D$ {\bf79}, 114018 (2009).
\bibitem{bc7}
 W. Wang, Y. L. Shen and C. D. Lu, $Phys.~Rev.~D$ {\bf79}, 054012 (2009).
\bibitem{bc8}
 X. Liu and Z. J. Xiao, $Phys.~Rev.~D$ {\bf81}, 074017 (2010).

\bibitem{bc-pwave}
Z. H. Wang, G. L. Wang and C. H. Chang, $J.~Phys.~G$ {\bf39}, 015009(2012).
\bibitem{heavy-light}
Z. H. Wang, G. L. Wang, H. F. Fu and Y. Jiang, $Int.~J.~Mod.~Phys.~A$ {\bf27}, 1250049(2012).
\bibitem{bc9}
H. F. fu, Y. Jiang, C. S. Kim and G. L. Wang, $JHEP$ {\bf1106}, 015(2011).
\bibitem{bc10}
W. L. Ju, G. L. Wang, H. F. Fu, T. H. Wang and Y, Jiang, $JHEP$ {\bf1404}, 065(2014).
\bibitem{bc11}
W. L. Ju, G. L. Wang, H. F. Fu, Z. H. Wang and Y. Li, $JHEP$ {\bf1509}, 171(2015).
\bibitem{bc12}
W. L. Ju, T. H. Wang, Y. Jiang, H. Yuan and G. L. Wang, $J.~Phys.~G$ {\bf43}, 045004(2016).
\bibitem{bc13}
C. H. Chang, H. F. Fu, G. L. Wang and J. M. Zhang, $Sci.~Chin.~Phys.~Mech.~Astron.$ {\bf58}, 071001(2015).
\bibitem{lhc1}
N. Brambilla $et al$., CERN Yellow Report, CERN-2005-005, and references therein.
\bibitem{lhc2}
N. Brambilla $et al.$, $Eur.~Phys.~J.~C$ {\bf71}, 1534 (2011).
\bibitem{robert1}
P. Maris and C. D. Roberts, $Phys.~Rev.~C$ {\bf56}, 3369 (1997).
\bibitem{robert2}
P. Maris, C. D. Roberts and P. C. Tandy, $Phys.~Lett.~B$ {\bf420}, 267 (1998).
\bibitem{robert3}
M. A. Ivanov, Yu. L. Kalinovsky, P. Maris and C. D. Roberts, $Phys.~Rev.~C$ {\bf57} 1991(1998).
\bibitem{robert4}B. El-Bennich, M. A. Ivanov and C. D. Roberts, $Phys.~Rev.~C$ {\bf83} 025205 (2011).
\bibitem{BS1}
C. H. Chang, J. K. Chen and G. L. Wang, $Commun.~Theor.~Phys.$ {\bf46}, 467(2006).
\bibitem{Heff1}
G. Buchalla, A. J. Buras and M. E. Lautenbacher, $Rev.~Mod.~Phys.$ {\bf68}, 1125(1996).
\bibitem{Heff2}
A. Ali, J. Chay. C. Greub and P.ko, $Phys.~Lett.~B$ {\bf424}, 161(1998).
\bibitem{naive}
M. Bauer, B. Stech and M. Wirbel, $Z.~Phys.~C$ {\bf34}, 103(1987).
\bibitem{pseudo}
G. Cvetic, C. S. Kim, G. L. Wang and Wuk Namgung, $Phys.~Lett.~B$, {\bf596}, 84(2004).
\bibitem{vector}
G.L. Wang, $Phys.~Lett.~B$, {\bf633}, 492(2006).
\bibitem{Mand} S. Mandelstam, $Proc.~R.~Soc.~London$ {\bf 233}, 248(1955).

\bibitem{w1} C. S. Kim, G. L. Wang, $Phys.~Lett.~B$ {\bf584}, 285(2004).
\bibitem{mass1}
C. H. Chang and G. L. Wang, $Science~in~China~Series~G$
{\bf53}, 2005(2010).
\bibitem{kk}
P. Ball and R. Zwicky, $Phys.~Rev.~D$ {\bf71}, 014029 (2005).
\bibitem{BS} E.E. Salpeter and H.A. Bethe, $Phys.~Rev.$ {\bf84}, 1232(1951).
\bibitem{eff1}
C. F. Qiao, H. W. Huang and K. T Zhao,  $Phys.~Rev.~D$ {\bf54}, 2273 (1996).
\bibitem{eff2}
C. F. Qiao, H. W. Huang and K. T Zhao,  $Phys.~Rev.~D$ {\bf60}, 094004 (1999).
\bibitem{eff3}
D. Ebert, R. N. Faustov and V. O. Galkin, $Phys.~Rev.~D$ {\bf62}, 034014 (2000).
\bibitem{Salp}  E.E. Salpeter, $Phys.~Rev.$ {\bf87}, 328(1952).



\end{thebibliography}
\end{document}